\documentclass{aastex}

\received{}
\accepted{}

\slugcomment{Submitted to ApJ Letters}

\newcommand{\ms}{\mbox{m s$^{-1}$}}
\newcommand{\msun}{M$_{\odot}$}

\newcommand{\mjup}{M$_{\rm JUP}$}
\newcommand{\msat}{M$_{\rm SAT}$}

\newcommand{\msini}{$M \sin i~$}

\shortauthors{Marcy {\it et~al.\/}}
\shorttitle{Two New Planets}
\begin{document}

\title{Sub-Saturn Planet Candidates to HD 16141 and HD 46375$~^{1}$}

\author{Geoffrey W. Marcy\altaffilmark{2},
R. Paul Butler\altaffilmark{3}, 
Steven S. Vogt\altaffilmark{4}} 

\email{gmarcy@etoile.berkeley.edu}

\altaffiltext{1}{Based on observations obtained at the
W.M. Keck Observatory, which is operated jointly by the
University of California and the California Institute of Technology.}

\altaffiltext{2}{ Department of Astronomy, University of California,
Berkeley, CA USA 94720 and at Department of Physics and Astronomy,
San Francisco, CA, USA 94132}

\altaffiltext{4}{Department of Terrestrial Magnetism, Carnegie Institution
of Washington, 5241 Broad Branch Road NW, Washington D.C., USA 20015-1305}

\altaffiltext{4}{UCO/Lick Observatory, University of California at Santa Cruz,
Santa Cruz, CA, USA 95064}

\begin{abstract}

Precision Doppler measurements from the Keck/HIRES spectrometer reveal
periodic Keplerian velocity variations in the stars HD 16141 and HD
46375.  HD 16141 (G5 IV) has a period of 75.8 d and a velocity
amplitude of 11 \ms, yielding a companion having \msini = 0.22 \mjup \
and a semimajor axis, $a$ = 0.35 AU.  HD 46375 (K1 IV/V) has a period
of 3.024 d and a velocity amplitude of 35 \ms, yielding a companion
with \msini=0.25 \mjup, a semimajor axis of $a$ = 0.041 AU, and an
eccentricity of 0.04 (consistent with zero).  These companions contribute to the
rising planet mass function toward lower masses.

\end{abstract}


\section{Introduction}
\label{intro}

The two thousand nearest and brightest dwarf stars ranging in spectral
type from late F through M are currently being surveyed by groups with
a Doppler precision of $\sim$10 \ms \ or better.  To date
these surveys have resulted in the discoveries of 32 extrasolar
planets (cf. Marcy, Cochran, and Mayor 2000, Marcy \& Butler 2000, Vogt et
al. 2000, Queloz et al. (2000), Udry et al. (2000), Noyes et al. 1997), including the first
system of multiple planets (Butler et al. 1999) and the first
detection of a transiting planet (Henry et al. 2000; Charbonneau et
al. 2000).

Remarkably, all 32 companions found from precision Doppler surveys
have \msini less than 8 \mjup, though companions with masses of 10--80
\mjup \ would have been much easier to detect.  The companion mass
function rises steeply toward smaller masses (Marcy \& Butler 2000),
but it turns over near 0.5 \mjup, presumably due to poor
detectability.  With conventional precision of 10 \ms, companions of
0.5 \mjup \ in 4-day orbits induce stellar motion only a few times
greater than such Doppler errors.  The stars 51 Peg and HD 75289 have
the lowest known values of \msini, both 0.46 \mjup \ (Mayor \& Queloz
1995, Udry et al. 2000).  However with a precision of 3 \ms, one may
explore the mass distribution below 1 \msat \ (= 0.298 \mjup).  If
sub-Saturn mass companions occur less frequently than Jupiter-mass
companions, the status of all Jupiter-mass companions as ``planets''
would be cast in doubt.  Such a distribution of masses, peaked at
$\sim$1 \mjup, is not the case in our Solar System nor do theories of
planet formation predict such a peak (eg. Lissauer 1995, Boss 1995,
Levison et al. 1998).

Here we report Doppler variations in HD 16141 and HD 46375 exhibiting
\msini $<$ 1 \msat.  Stellar characteristics and observations are
discussed in the second section.  Orbital solutions are presented in
the third section, followed by discussion of results.

\section{Observations and Stellar Characteristics}
\label{obs}

HD 16141 (79 Cet, HIP 12048, G5 IV) and HD 46375 (HIP 31246, K1 IV-V)
are slowly rotating, chromospherically inactive stars, based on the
weak Ca II H\&K emission in their spectra.  Their Hipparcos (Perryman
et al. 1997) distances are 35.9 and 33.4 pc, respectively, giving them
absolute magnitudes of M$_V$=4.05 and M$_V$= 5.29, which places both
stars 1.0 mag above the zero-age main sequence.  The Hipparcos mission
made 60 and 63 photometric observations of HD 16141 and HD 46375,
respectively, revealing that both stars are photometrically stable at
the level of errors, $\sim$0.01 mag.

The mass of HD 16141 is $M$=1.01 \msun, and it has [Fe/H] = 0.02, as
determined by high-resolution spectral synthesis (Fuhrmann 1998). The
mass of HD 46375 is probably $\sim$1.0 \msun also.  Its spectral type of
K1 IV-V and color, $B-V$= 0.86, would normally indicate a mass of 0.9
\msun.  But its metalicity appears to be high, [Fe/H]=$\sim$+0.34, as
judged from narrow band photometry (Apps, priv. comm. 2000).  This
high metalicity implies a higher mass than is associated with
solar-metalicity stars of its color, giving it an estimated mass of
0.9 - 1.0 \msun.  Here we adopt a mass of 1.0$\pm$0.1 \msun, for
computation of \msini and semimajor axis.  HD 46375 resides only a few
arcminutes North-East from the bright nebulosity of the Rosette Nebula
(which is over 1 kpc in the background), leaving some concern about
contamination of the optical spectra and photometry of the star.

The velocities of HD 16141 and HD 46375 have been monitored since 1996
and 1998 respectively.  The technique is the same as described in Vogt
et al.  (2000), using the Keck 1 HIRES spectrometer (Vogt et
al. 1994).  The H\&K lines near 3950 \AA \ provide a simultaneous
chromospheric diagnostic (Saar et al. 1998), giving log$R$'(HK) =
$-$5.05 and $-$4.94 for HD 16141 and HD 46375, respectively, typical
for chromospherically quiet stars (Noyes et al. 1984).  The
photospheric Doppler ``jitter'' of such stars is less than 3 \ms (Saar
et al. 1998).
\section{Orbital Solutions}

The 46 observations of HD 16141 are shown in Figure 1 and listed in
Table 1.  These observations have an RMS of 7.0 \ms, 2.5 times larger
than the measurement error.  A periodogram of these velocities (Figure
2a), reveals a dominant period at 75.6 days.  The reality of this
periodicity is supported by two tests.  We broke the velocity set into
its first half and second half, yielding separate periodograms with highest
peaks at 75 d and 77 d, respectively, both having a false alarm
probability under 1\%.  Thus, both halves of the measurements reveal
the 76-day periodicity.  We also computed the false alarm probability
of the 76 d periodicity in the full velocity set by using a
Monte Carlo approach (Gilliland \& Baliunas 1987).  We generated
10$^5$ sets of artificial velocities drawn from a Gaussian error
distribution while retaining the actual times of
observation.  None of these artificial velocity sets yielded a
periodogram peak as high as that actually found for HD 16141 (Fig 2a).
Thus, the false alarm probability is less than
1$\times$10$^{-5}$.

The best--fit Keplerian model to the velocities of HD 16141 is shown
in Figure 1, and yields a period, $P$=75.82 d.  The amplitude is,
$K$=10.8 \ms, and the eccentricity is $e$=0.28.  The RMS of the
residuals to the Keplerian fit is 3.2 \ms, similar to the median
internal error of 2.8 \ms.  Adopting a stellar mass of 1 \msun, the
companion has \msini=0.22 \mjup, and the semimajor axis is 0.35 AU.  A
periodogram of the velocity residuals is shown in Figure 2b which
shows no significant peaks, indicating that no additional companions
are evident.

The 24 observations of HD 46375 are listed in Table 2.  Figure 3 shows
observations obtained between 6 and 11 Feb 2000, revealing a three day
period.  A phased version of the entire data set spanning 515 d is
shown in Figure 4.  The same best--fit sinusoid is shown in both
Figures 3 and 4.  The RMS of the sinusoidal fit is 2.59 \ms, slightly
greater than the velocity uncertainty of 2.2 \ms.  The best--fit
Keplerian gives $e$=0.04$\pm$0.04 (i.e. not significant), and yields
an RMS of 2.44 \ms.  The period is 3.024 d and the velocity
semiamplitude $K$ = 35 \ms.  Assuming the host star is 1 \msun, the
companion has \msini=0.25 \mjup \ and the semimajor axis is 0.041 AU.
The orbital parameters for HD 16141 and HD 46375 are listed in Table 3.

\section{Discussion}

HD 16141 exhibits the smallest velocity amplitude, $K$=10.8 \ms,
reported for any planet candidate, thus warranting an assessment of
stellar and instrumental sources of error.  Chromospherically inactive
stars are intrinsically stable at the level of 3 \ms (Saar et
al. 1998).  Further, the majority of our program stars exhibit no
instrumental ``drift'' in the velocity zero--point at the 2 m/s level
during the last four years (Vogt et al. 2000).  There is no plausible
stellar time scale near 75 d except rotation.  But spots are not
important as HD 16141 is chromospherically quiet at Ca II H\&K, and
photometrically stable at 0.01 mag from Hipparcos.  Thus, the most
likely explanation for the Doppler variations is an orbiting planet.

The eccentricity for the planet around HD 16141 is not yet well
determined, $e$=0.28$\pm$0.15, and indeed cannot be reliably
distinguished from circular without further measurements.  However,
all 21 planets orbiting beyond 0.2 AU (Marcy \& Butler 2000) have
eccentricities above 0.1, and thus this planet will constitute an
interesting test case of eccentricities of a possibly low mass planet.

The companion to HD 46375 is similar to the other ``51 Peg--like''
planets with their orbital periods of 3 to 5 days and circular orbits.
Circular orbits are expected from tidal coupling with the primary star
(Rasio \& Ford 1996; Marcy et al. 1997; Ford et al. 1998).  As the
primary in the HD 46375 system has not been spun up, the mass of the
planet is constrained to be less than $\sim$15 \mjup (Ford et al. 1998, Marcy
et al. 1997).  The suspected high metalicity of HD 46375 of [Fe/H]=0.34
(Apps, priv. comm.)  supports the suggestion that ``51 Peg--like''planets are 
associated with high-metalicity stars (Gonzalez et al. 1999,
Queloz et al. 2000) .

The value of \msini=0.25 \mjup \ for the companion to HD 46375 is
smaller than that of any known ``51 Peg--like'' extrasolar planet,
with the smallest having been 51 Peg (0.46 \mjup, Mayor \& Queloz
1995) and HD 75289 (0.46 \mjup, Udry et al. 2000).  This suggests that
if orbital migration brings such planets inward (Lin et al. 1996), the
process may continue to operate at masses near 1 \msat, pending knowledge of
$\sin i$.

With demonstrated precision of 3 \ms (Vogt et al. 2000), the Keck
survey is currently capable of making 3 $\sigma$ detections of ``51
Peg--like'' planets down to the neptune--mass range.  Knowledge of the
companion mass function of ``51 Peg--like'' planets down into this
range will provide useful constraints on models that explain the
formation and subsequent dynamics of ``51 Peg--like'' planets.

The effective temperature of a planet in a 3.024 day orbit about HD
46375 would be $\sim$1400 K (Burrows et al. 1998).  Henry (2000) reports
that no transit occurs, implying that $\sin i <$ 0.992 .
The expected amplitude of a
transit signal is about 0.015 mags (Burrows et al. 1998, Henry et
al. 2000, Charbonneau et al. 2000).

The companions to HD 16141 and HD 46375 have the lowest values of
\msini (0.22 \mjup, 0.25 \mjup) found to date for extrasolar planets.
The observed histogram of \msini shows a steep rise toward the lowest
masses, consistent with a power-law, d$N$/d$M$ $\propto M^{-1}$ (Marcy
and Butler 2000).  These new planets support the
suggestion that the mass distribution continues rising
to 1 \msat.  Verification of
any rise in the planetary mass function below 1 \msat will
require more detections to account properly for incompleteness.

\acknowledgements

We thank Kevin Apps for assessment of stellar characteristics.
We acknowledge support by NASA grant NAG5-8299 and NSF grant
AST95-20443 (to GWM), by NSF grant AST-9619418 and NASA grant
NAG5-4445 (to SSV), travel support from the Carnegie Institution
of Washington (to RPB), and by Sun Microsystems. 
This research has made use of the SIMBAD database,
operated at CDS, Strasbourg, France.

\clearpage
\begin{figure}
%\plotone{fig1.ps}
\caption{Keck Doppler velocities, with error bars, for HD 16141.
The solid line through the points shows the best Keplerian fit.  
The residuals having rms=3.24 \ms are shown at bottom
(crosses) with arbitrary zero-point (horizontal line).}

\label{rv_curve}
\end{figure}

\begin{figure}
%\plotone{fig2.ps}
\caption{a) Periodogram of the measured Doppler velocities of HD 16141.
The dotted line shows the 1\% false alarm level.  The highest
peak at 75.6 days has a false alarm probability less than 1$\times$10$^{-5}$
from Monte Carlo tests.
b)  Periodogram of the velocities of HD 16141, after removing the
best fit Keplerian.  No significant peaks remain.}
\label{fig2}
\end{figure}

%\begin{figure}
%\plotone{fig3.ps}
%\caption{Phased velocities of HD 16141 using the period from 
%the best--fit Keplerian.  The Keplerian is shown as the solid
%line.  The period is 75.82 d, the semiamplitude is 10.7 \ms,
%and the eccentricity is 0.28.  Assuming HD 16141 is 1 \msun,
%the minimum (\msini) mass of the companion is 0.22 \mjup, and
%the semimajor axis is 0.35 AU.}
%\label{fig3}
%\end{figure}

\begin{figure}
%\plotone{fig3.ps}
\caption{Keck Doppler velocities for HD 46375 from the 2000 Feb
observing run.  A 3 day periodicity is evident for this 6 night
observing string.  Measurement uncertainties are $\sim$ 2.2 \ms.}
\label{fig3}
\end{figure}

\begin{figure}
%\plotone{fig4.ps}
\caption{Phased Doppler velocities for HD 46375.  The same best--fit
sinusoid is shown in Figures 4 and 5.  The RMS to the sinusoidal fit
is 2.59 \ms, essentially identical to the best--fit Keplerian.  The period
is 3.024 d and the semiamplitude is 35 \ms.  Assuming the mass of HD 46375
is 1 \msun, the minimum (\msini) mass of the companion is 0.25 \mjup,
and the semimajor axis is 0.041 AU.}
\label{fig4}
\end{figure}

\clearpage

\begin{deluxetable}{rrr}
\tablenum{1}
\tablecaption{Velocities for HD 16141}
\label{vel16141}
\tablewidth{0pt}
\tablehead{
JD & RV & error \\
(-2450000)   &  (m s$^{-1}$) & (m s$^{-1}$)
}
\startdata
\tableline
   365.9888  & -15.1  &  2.5 \\
   715.1077  &  10.9  &  2.7 \\
   786.8339  &  16.9  &  3.1 \\
   806.8945  & -10.4  &  2.9 \\
   837.7389  &  -3.5  &  2.5 \\
   838.7150  &   1.4  &  2.5 \\
   839.7379  &   0.0  &  2.5 \\
  1010.1177  &  10.6  &  2.5 \\
  1013.1171  &  15.6  &  2.9 \\
  1043.0747  &  -9.6  &  3.0 \\
  1044.1033  &  -8.5  &  2.5 \\
  1050.9976  &  -6.2  &  2.8 \\
  1068.9550  &   2.8  &  2.2 \\
  1070.1031  &  -0.3  &  2.4 \\
  1070.9812  &   8.2  &  3.0 \\
  1071.9963  &   2.4  &  3.1 \\
  1072.9689  &   1.5  &  2.7 \\
  1074.9096  &   4.2  &  2.8 \\
  1170.8101  &   1.3  &  2.7 \\
  1171.7517  &   0.7  &  2.6 \\
  1172.8463  &  -1.8  &  2.8 \\
  1173.7982  &   0.0  &  2.9 \\
  1226.7259  &   4.4  &  2.4 \\
  1227.7267  &   0.2  &  2.0 \\
  1228.7149  &   6.0  &  2.5 \\
  1368.1243  &   5.8  &  3.3 \\
  1370.1264  &  -2.1  &  3.0 \\
  1371.1179  &   2.4  &  2.7 \\
  1373.1326  &   5.3  &  2.7 \\
  1374.1180  &   4.0  &  3.1 \\
  1410.1390  &  -3.9  &  2.8 \\
  1411.0372  &  -2.5  &  3.0 \\
  1412.0972  &  -5.5  &  3.3 \\
  1438.9326  &  -3.2  &  3.1 \\
  1439.9281  &  -0.6  &  3.0 \\
  1440.9359  &  -5.0  &  3.0 \\
  1441.9363  &  -4.4  &  2.8 \\
  1543.8553  &  12.4  &  2.8 \\
  1550.7971  &   4.1  &  3.0 \\
  1551.7998  &   5.4  &  2.5 \\
  1552.8500  &   7.9  &  2.8 \\
  1580.7274  &  -7.0  &  2.8 \\
  1581.7594  &  -8.9  &  2.1 \\
  1582.7209  &  -9.9  &  2.0 \\
  1583.7356  &  -7.4  &  2.1 \\
  1585.7334  &  -2.2  &  2.9 \\
\enddata
\end{deluxetable}

\clearpage

\begin{deluxetable}{rrr}
\tablenum{2}
\tablecaption{Velocities for HD 46375}
\label{vel46375}
\tablewidth{0pt}
\tablehead{
JD & RV & error \\
(-2450000)   &  (m s$^{-1}$) & (m s$^{-1}$)
}
\startdata
\tableline

  1070.1367  &  16.4  &  2.5 \\
  1171.9009  & -38.0  &  2.5 \\
  1226.8575  & -20.2  &  3.7 \\
  1227.8334  &  30.3  &  2.6 \\
  1228.9406  & -37.2  &  3.7 \\
  1544.0196  & -35.2  &  2.6 \\
  1550.9756  &  22.3  &  2.0 \\
  1551.7716  &  17.3  &  2.2 \\
  1551.9140  &   5.1  &  1.7 \\
  1552.0074  &   2.3  &  1.7 \\
  1552.8292  & -37.3  &  2.1 \\
  1552.9381  & -39.5  &  2.3 \\
  1580.8203  &  -5.8  &  2.0 \\
  1580.9892  &   0.5  &  1.4 \\
  1581.7296  &  25.3  &  2.4 \\
  1581.9588  &  16.8  &  2.3 \\
  1582.7138  & -33.5  &  1.7 \\
  1582.7945  & -34.2  &  2.1 \\
  1582.8802  & -39.2  &  1.4 \\
  1583.7291  & -16.9  &  1.3 \\
  1583.9223  &   0.2  &  2.2 \\
  1584.8240  &  23.0  &  3.5 \\
  1585.8850  & -41.6  &  1.4 \\
  1585.9671  & -42.2  &  1.9 \\
\enddata
\end{deluxetable}

\clearpage

\begin{deluxetable}{lcc}
\tablenum{3}
\tablecaption{Orbital Parameters}
\label{orbit}
\tablewidth{0pt}
\tablehead{
\colhead{Parameter} & \colhead{HD 16141} & \colhead{HD 46375}
}
\startdata
Orbital period $P$ (d) &  75.82 (0.4) & 3.024 (0.0005) \\
Velocity amp. $K$ (m\,s$^{-1}$) & 10.8 (1.4) & 35.2 (1.7) \\
Eccentricity $e$  & 0.28 (0.15) & 0.04 (0.04) \\
$\omega$ (deg)    & 41 (22)     & 62 (50) \\
Periastron Time (JD) & 2451547.2 (4.1) & 2451582.10 (0.23) \\
Msini (\mjup)     &  0.215 (0.03)       & 0.249 (0.03) \\
a (AU)            &  0.35       & 0.041  \\
\enddata
\end{deluxetable}
 
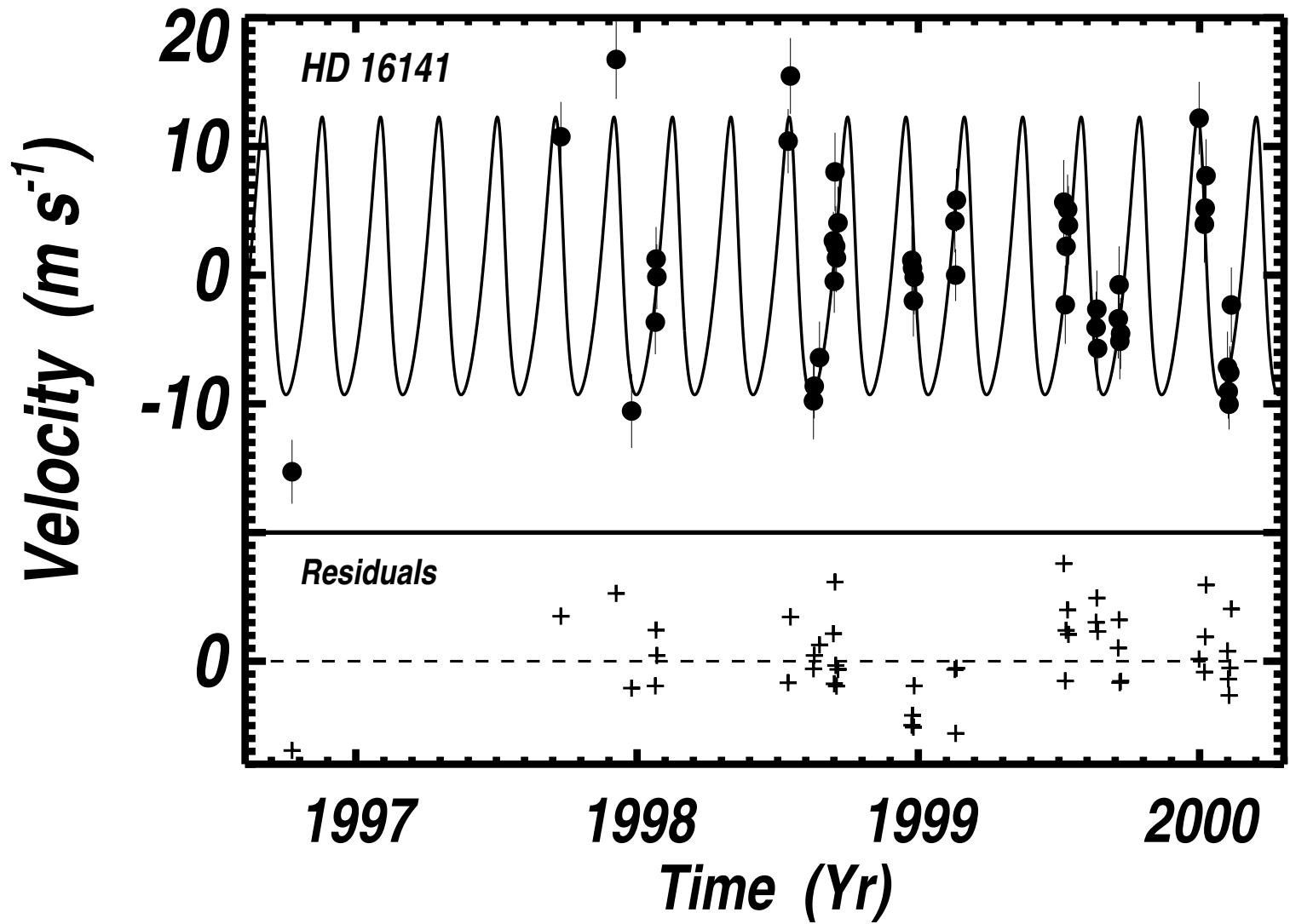
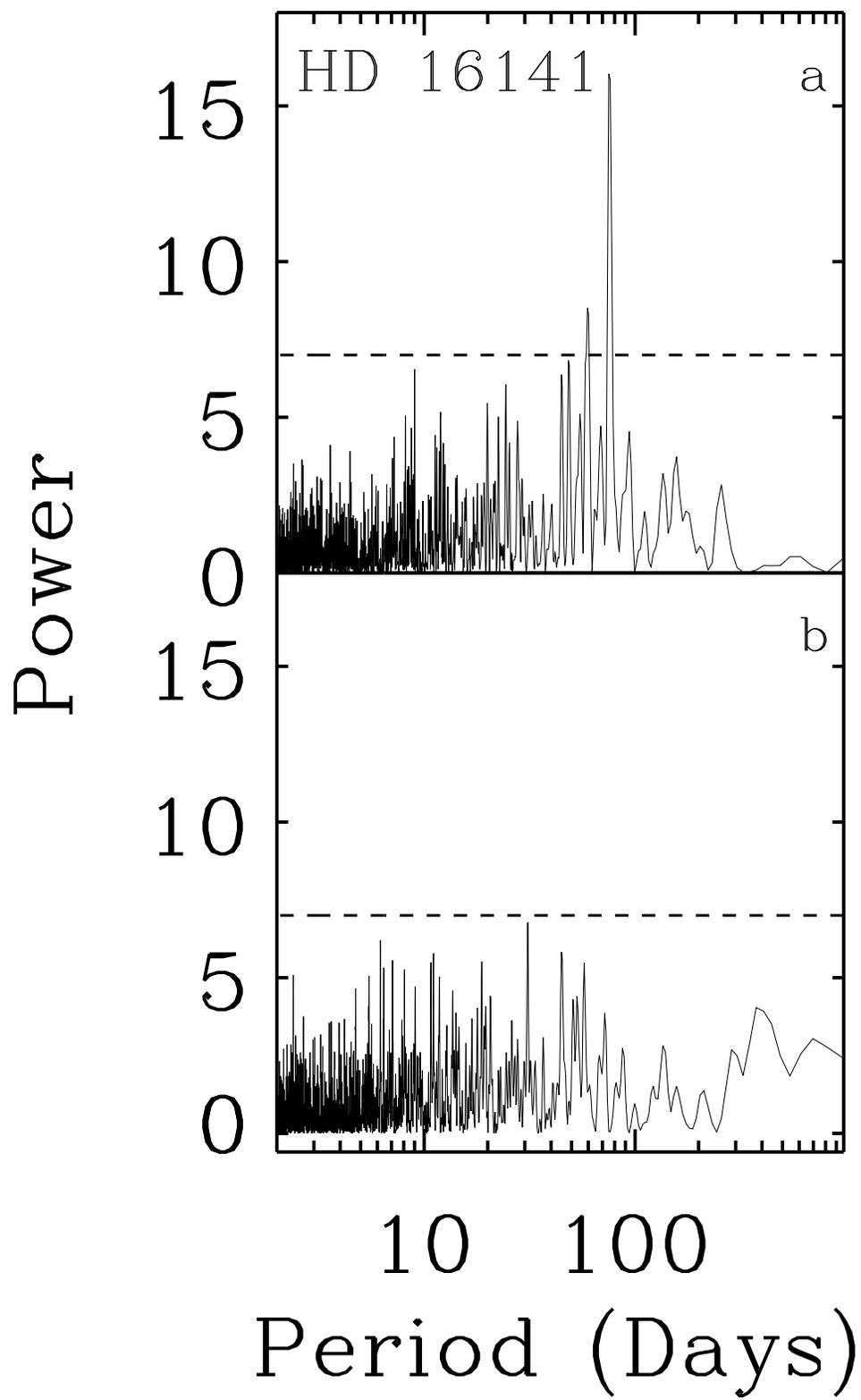
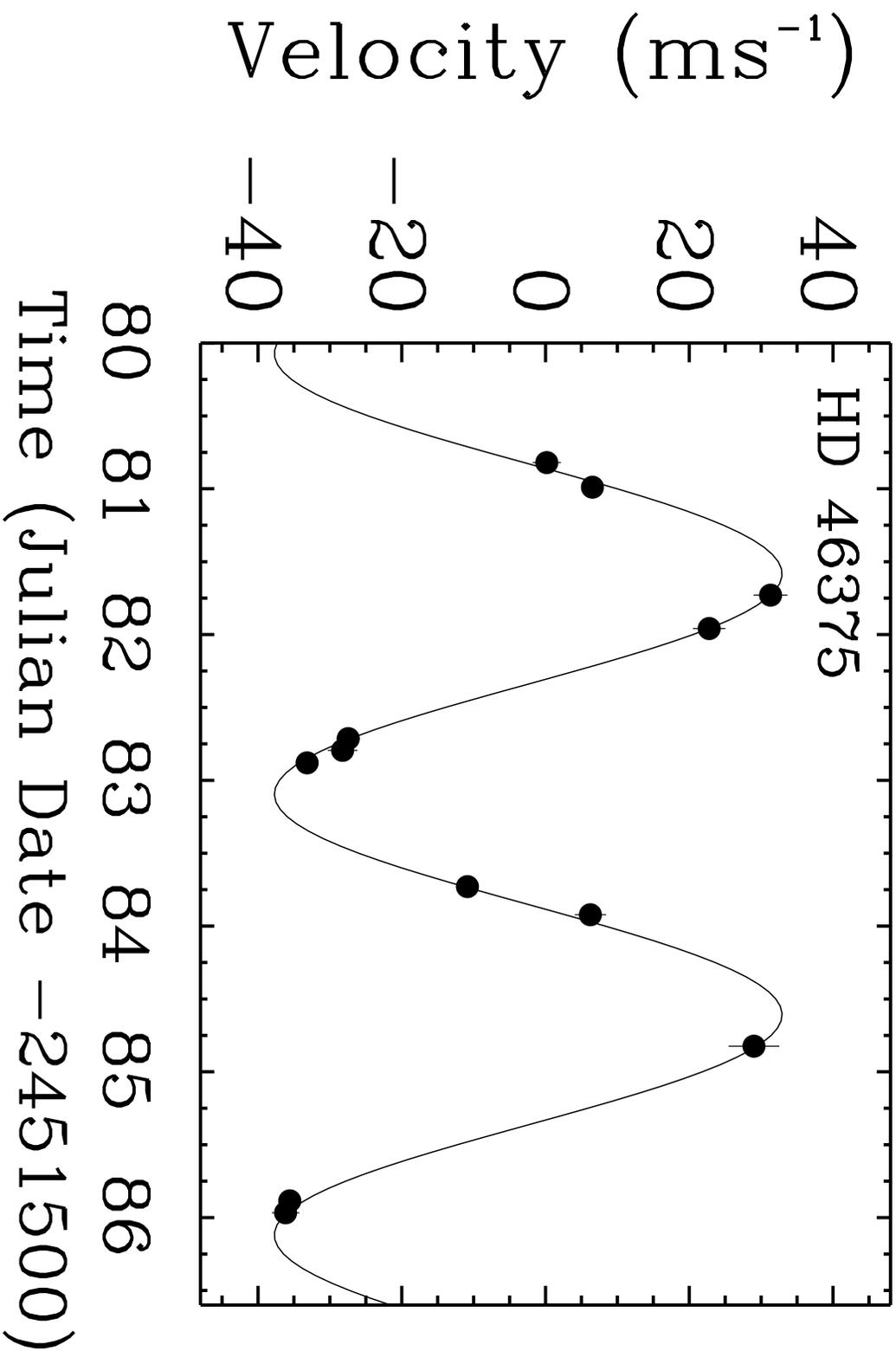
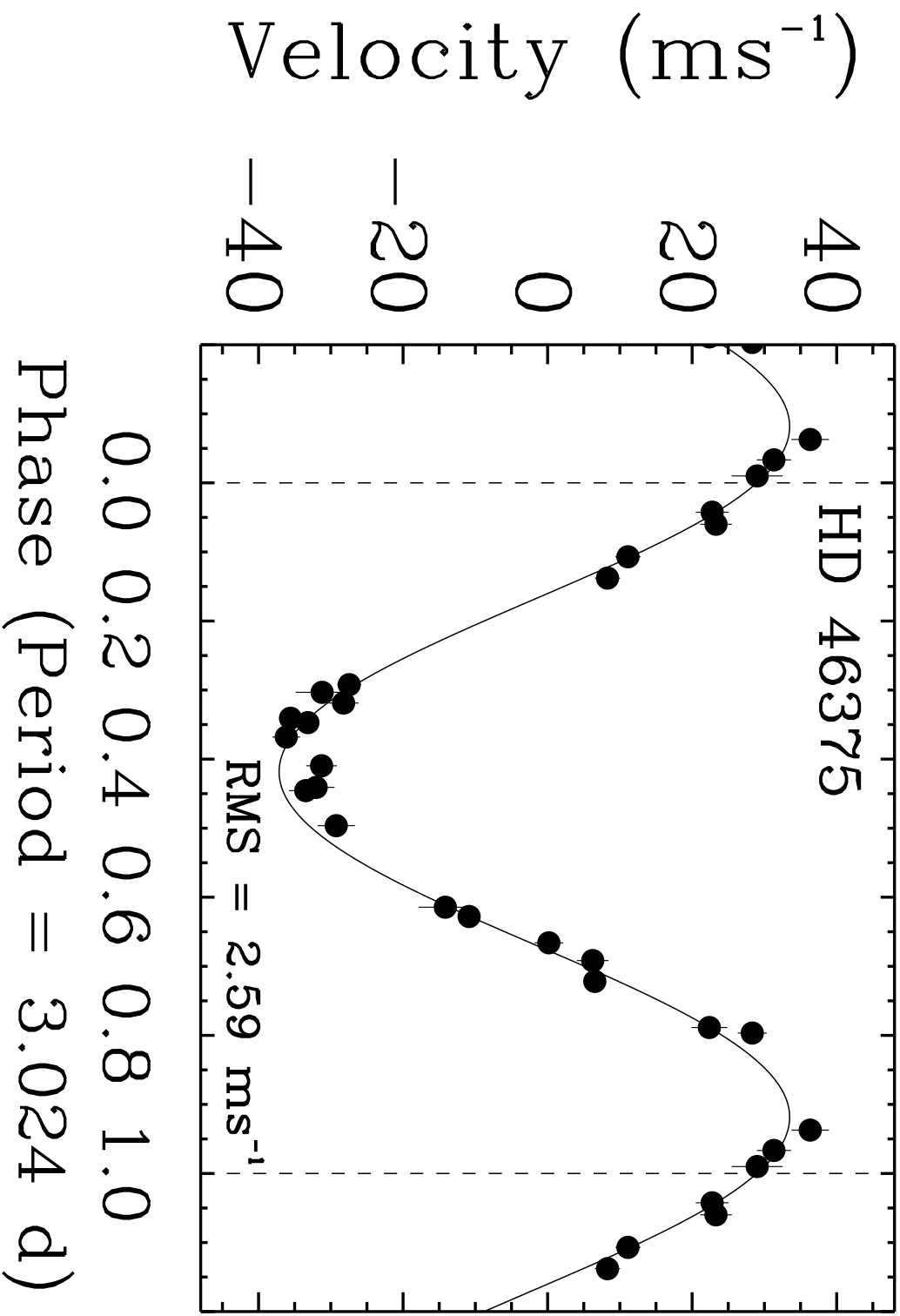
\end{document}